\documentclass[12pt]{article}

\usepackage{amssymb,amsfonts,amsmath,bm,color}
\usepackage{graphicx}

\begin{document}

\vskip 2cm

\begin{center}
{\Large\bf Expanding Polyhedral Universe in Regge Calculus}
\end{center}
\vspace*{1cm}
\begin{center}{\sc Ren Tsuda${}^1$ and Takanori Fujiwara${}^2$}
\end{center}
\vspace*{0.2cm}
\begin{center}
{\it ${}^1$ Graduate School of Science and Engineering, Ibaraki University,
Mito 310-8512, Japan \\
${}^2$ Department of Physics, Ibaraki University,
Mito 310-8512, Japan}
\end{center}

\vfill

%%%%%%%%%%%%%%%%%%%%%%%%%%%%%%%%%%%%%%%%%%%%
%%%%%%%%%%%%%%%%%%%%%%%%%%%%%%%%%%%%%%%%%%%%
%%%%%%%%%%%%%%%%%%%%%%%%%%%%%%%%%%%%%%%%%%%%

%   << abstract >>

%%%%%%%%%%%%%%%%%%%%%%%%%%%%%%%%%%%%%%%%%%%%
%%%%%%%%%%%%%%%%%%%%%%%%%%%%%%%%%%%%%%%%%%%%
%%%%%%%%%%%%%%%%%%%%%%%%%%%%%%%%%%%%%%%%%%%%

\begin{abstract} 
The closed Friedmann--Lema\^itre--Robertson--Walker (FLRW) universe of Einstein gravity 
with positive cosmological constant in three dimensions is investigated 
by using the Collins--Williams formalism in Regge calculus. 
A spherical Cauchy surface is replaced with regular polyhedrons. 
The Regge equations are reduced to differential equations 
in the continuum time limit. Numerical solutions to the Regge equations 
approximate well the continuum FLRW universe during the era of 
small edge length. The deviation from the continuum solution 
becomes larger and larger with time. Unlike 
the continuum universe, the polyhedral universe expands to 
infinite within finite time. To remedy the shortcoming 
of the model universe we introduce geodesic domes and 
pseudo-regular polyhedrons. 
It is shown that the pseudo-regular polyhedron 
model can approximate well the results of the Regge calculus for the 
geodesic domes. The pseudo-regular polyhedron model approaches 
the continuum solution in the infinite frequency limit. 
\end{abstract}

\newpage

\pagestyle{plain}

%%%%%%%%%%%%%%%%%%%%%%%%%%%%%%%%%%%%%%%%%%%%
%%%%%%%%%%%%%%%%%%%%%%%%%%%%%%%%%%%%%%%%%%%%
%%%%%%%%%%%%%%%%%%%%%%%%%%%%%%%%%%%%%%%%%%%%

%        << Sec.1 Introduction >>

%%%%%%%%%%%%%%%%%%%%%%%%%%%%%%%%%%%%%%%%%%%%
%%%%%%%%%%%%%%%%%%%%%%%%%%%%%%%%%%%%%%%%%%%%
%%%%%%%%%%%%%%%%%%%%%%%%%%%%%%%%%%%%%%%%%%%%

\section{Introduction}

\label{sec:intro}
\setcounter{equation}{0}

Regge calculus was proposed to formulate Einstein's general relativity 
on piecewise linear manifolds \cite{Regge:1961aa, MTW}. 
It provides a coordinate-free lattice formulation of gravitation
and has been used in investigations of classical as well as quantum 
gravity. Like QCD, the lattice theoretical approach 
provides a powerful framework in nonperturbative studies 
of quantum gravity \cite{David:1992aa}. However, before moving to detailed quantum 
study, the formalism at the classical level should be investigated. 
Any lattice regularized theory should reproduce the basic 
results of the corresponding continuum theory. Taking the classical continuum 
limit is relatively easy in the case of lattice gauge theories. The 
reason for this is obvious: In lattice gauge theory, space-time 
itself is not dynamical. We usually consider a hypercubic regular 
lattice for the space-time. Dynamical variables sit on sites for 
matter fields and on links for gauge fields. Classical lattice actions are 
written in manifestly gauge-invariant form by using plaquette 
variables and covariant differences, which have obvious classical 
counterparts. 
The point is that in the lattice gauge it is easy, at least classically, 
to investigate how the theories behave under changes of lattice size 
and lattice spacing.

In Regge calculus, the space-time is replaced with a piecewise 
linear manifold, which is composed of a set of simplices. The 
basic variables are the edge lengths. As in general relativity, 
the space-time itself should be considered dynamical. We do not 
know, however, the space-time to be investigated precisely from 
the beginning. To prepare the Regge action we must assume the 
topology of the space-time. For a given topology we can triangulate 
the space-time and write the Regge action. In general there is 
no natural choice of the piecewise linear manifold. Furthermore, 
the Regge action is written in coordinate-free form. It is a 
highly complicated function of the edge length depending heavily 
on the triangulations of the space-time. This makes 
investigations of how the theory behaves with respect to
refinement of the triangulation much more involved than the 
lattice gauge theory. 

In this note we investigate the Friedmann--Lema\^itre--Robertson--Walker (FLRW) 
universe of three-dimensional Einstein gravity with positive cosmological 
constant in Regge calculus by taking polyhedrons as the 
Cauchy surface. We compare the solutions 
between regular polyhedrons, and propose a generalization of 
the Regge equations beyond them; this makes the numerical 
analysis much easier than the orthodox Regge calculus.  

The continuum action is given by
\begin{eqnarray}
  \label{eq:cEHa}
  S=\frac{1}{16\pi}\int d^3x\sqrt{-g}(R-2\Lambda).
\end{eqnarray}
It is well known in three dimensions that the vacuum Einstein equation 
leads to a flat space-time without the cosmological term. In the case of a negative 
cosmological constant the theory admits a black hole solution \cite{BTZ:1992aa,BHTZ:1993aa} 
and has been investigated within the context of conformal field 
theory \cite{Carlip:2005aa}. 
As in four dimensions, the Einstein equations have an evolving universe as 
a solution for the FLRW metric ansatz
\begin{eqnarray}
  \label{eq:FLRWm}
  ds^2=-dt^2+a(t)^2\left(\frac{dr^2}{1-kr^2}+r^2d\varphi^2\right),  
\end{eqnarray}
where $a(t)$ is the so-called scale 
factor. It is subject to the Friedmann equations
\begin{eqnarray}
  \label{eq:cc}
  \dot a^2=\Lambda a^2-k, \qquad
  \ddot a=\Lambda a. 
\end{eqnarray}
The curvature parameter $k=1,0,-1$ corresponds to space being spherical, 
Euclidean, or hyperspherical, respectively.  
Of these, our concern is the spherical 
universe, which can be approximated by a convex polyhedron of finite 
volume. Regge calculus has been applied to the four-dimensional closed 
FLRW universe by Collins and Williams \cite{CW:1973aa}. They 
considered regular polytopes as the Cauchy surfaces of the discrete 
FLRW universe and used, instead of simplices, truncated world-tubes 
evolving from one Cauchy surface to the next as the building blocks 
of piecewise linear space-time. Their method, called the Collins--Williams (CW) 
formalism, is based on the idea of the $3+1$ decomposition of 
space-time and plays a similar role to the well-known Arnowitt--Deser--Misner (ADM) 
formalism \cite{Brewin:1987aa}. Recently, Liu and Williams have 
extensively studied the discrete FLRW universe \cite{LW:2015aa,LW:2015ab}. 
They found that a universe with regular polytopes as the Cauchy surfaces can 
reproduce the continuum FLRW universe to a certain degree of 
precision. Their solutions agree reasonably well with the continuum when 
the size of the universe is small, whereas the deviations from 
the exact results become large for a large universe because 
of the finite edge length. Since the Regge action depends heavily 
on the choice of polytopes to approximate the Cauchy surface, 
it seems to be hard to take the continuum limit while keeping the action 
simple. This motivates us to investigate a simpler but less realistic 
three-dimensional model. 

In four dimensions there are six types of regular polytopes. 
Restricting to those obtained by tessellating the three-dimensional 
sphere by regular tetrahedrons, 
there are only three, with 5, 16, and 600 cells \cite{Coxeter}. 
The foregoing investigations are 
mainly restricted to these regular polytopes. The Regge equations, 
however, are still very complicated. As we shall show, the 
situation becomes much simpler in three dimensions, where 
every geometric calculation can be done without complications 
coming from higher dimensions. This is the reason why we consider 
three dimensions, where the spherical Cauchy surfaces are replaced
by regular polyhedrons. There are five types of polyhedrons. 
We treat them in a unified way and give generic expressions 
for the Regge equations that are convenient to analyze beyond the 
regular polyhedrons. 

Let us briefly summarize the essence of Regge calculus. 
In Regge calculus, an analog of the Einstein--Hilbert action is given 
by the Regge action \cite{Miller:1997aa}
\begin{eqnarray}
  \label{eq:ract}
  S_{\rm Regge}=\frac{1}{8\pi}\left(\sum_{i\in\rm \{hinges\}}
    \varepsilon_iA_i-\Lambda\sum_{i\in\rm \{blocks\}} V_i\right),
\end{eqnarray}
where $A_i$ is the volume of a hinge, $\varepsilon_i$ the deficit angle 
around the hinge $ A_i $, and $ V_i $ the volume of a 
building block of the piecewise linear manifold. In three dimensions 
the hinges are the links, or equivalently the edges of the 3-simplices, 
and $A_i$ is nothing but the edge length $l_i$. Regge's original 
derivation is concerned with a simplicial lattice, so that it 
describes the gravity as simplicial geometry. This formalism 
can easily be generalized to arbitrary lattice geometries. We can 
fully triangulate the non-simplicial flat blocks by adding extra 
hinges with vanishing deficit angles \cite{LW:2015aa} 
without affecting the Regge action.

The fundamental variables in Regge calculus are the edge lengths 
$l_i$. Varying the Regge action with respect to $l_i$, we obtain 
the Regge equations
\begin{eqnarray}
  \label{eq:regeq}
  \sum_{i\in \rm \{hinges\}}\varepsilon_i\frac{\partial A_i}{\partial l_j}
  -\Lambda\sum_{i\in \rm \{ blocks \}}\frac{\partial V_i}{\partial l_j}=0.
\end{eqnarray}
Note that there is no need to carry out the variation of the deficit angles 
owing to the Schl\"afli identity \cite{Schlafli:1858aa, HHKL:2015aa}
\begin{eqnarray}
\sum_{i \in \rm \{hinges\}} A_i \frac{\partial \varepsilon_i }{\partial l_j}=0.
\end{eqnarray}
In three dimensions, the Regge equations simply relate the 
deficit angle around an edge to the total rate of variation of 
the volumes having the edge in common with respect to the 
edge length. In particular, the space-time becomes flat in the 
absence of the cosmological term, as it should be. 

This paper is organized as follows: In the next section we set up 
the regular polyhedral universe in the CW formalism and introduce the Regge 
action. Derivation of the Regge equations is given in Sect. \ref{sec:req}. 
In the continuum time limit the Regge equations are reduced to differential 
equations. Applying the Wick rotation, we arrive at the Regge calculus analog of the Friedmann equations, describing the 
evolution of the polyhedral universe. This is done in Sect. \ref{sec:ctl}. 
In Sect. \ref{sec:nsol} we solve the differential Regge equation numerically 
and compare the scale factors of the polyhedral universe with the continuum 
solution. To obtain better approximations we introduce geodesic domes 
as the Cauchy surface. In Sect. \ref{sec:gdu} we propose a pseudo-regular 
polyhedral universe with a fractional Schl\"afli symbol  as a substitute 
for the geodesic dome universe and show that the features of the geodesic dome 
universe can be described well by the pseudo-regular polyhedron model. 
It is also argued that the continuum solution can be recovered in the 
infinite frequency limit. Section \ref{sec:sum} is devoted to summary and 
discussions. In Appendix \ref{sec:hcgd}, the Regge calculus for the first two, simplest, 
geodesic domes is described.

%%%%%%%%%%%%%%%%%%%%%%%%%%%%%%%%%%%%%%%%%%%%
%%%%%%%%%%%%%%%%%%%%%%%%%%%%%%%%%%%%%%%%%%%%
%%%%%%%%%%%%%%%%%%%%%%%%%%%%%%%%%%%%%%%%%%%%

%                                << Sec.2 Regge action for regular polyhedral universe >>

%%%%%%%%%%%%%%%%%%%%%%%%%%%%%%%%%%%%%%%%%%%%
%%%%%%%%%%%%%%%%%%%%%%%%%%%%%%%%%%%%%%%%%%%%
%%%%%%%%%%%%%%%%%%%%%%%%%%%%%%%%%%%%%%%%%%%%

\section{Regge action for a regular polyhedral universe}

\label{sec:ra}
\setcounter{equation}{0}

\begin{figure}[t]
  \centering
  \includegraphics[scale=1]{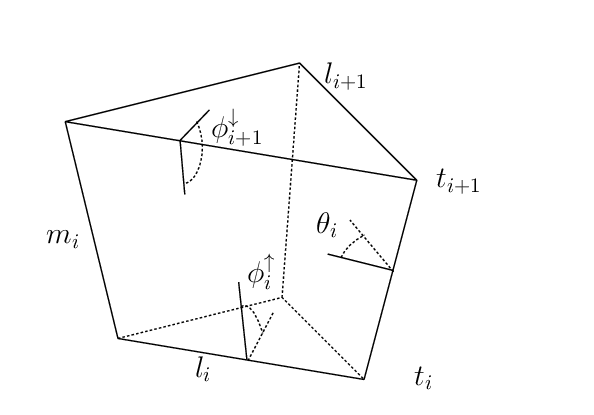}
  \caption{The $i$th frustum as the fundamental building block 
    of a polyhedral universe for $p=3$: 
    Each face of the regular polygon with edge length $l_i$ at time $t_i$ expands 
    to the upper one with $ l_{i+1} $ at $ t_{i+1} $.}
  \label{fig:fbb}
\end{figure}

The FLRW metric (\ref{eq:FLRWm}) describes an expanding or contracting 
universe with a maximally symmetric space as the Cauchy surface. Surfaces 
of maximally symmetric compact space are spheres, geometrically 
characterized by the radius $a(t)$, the scale factor in cosmology. 
In this paper we will be concerned with the Regge calculus of the closed 
FLRW universe in three dimensions, which describes an evolution of a 
two-dimensional sphere. Following the CW formalism, we replace the 
spherical Cauchy surfaces by regular polyhedrons. The fundamental 
building blocks of space-time are world-tubes of truncated pyramids or 
frustums, as depicted in Fig. \ref{fig:fbb}. In this section we 
restrict ourselves to regular polyhedrons as the Cauchy surfaces. Then 
every edge has equal length in each Cauchy surface, and so does any strut 
between two adjacent Cauchy surfaces. Evolution of the universe 
can be seen by focusing our attention only on expanding or shrinking of 
a face of the polyhedron. This considerably reduces the number of 
dynamical variables.

It is well known that there are only five types of regular polyhedron: 
tetrahedron, cube, octahedron, dodecahedron, and icosahedron. Let us 
denote the numbers of vertices, edges, and faces of a polyhedron by 
$N_0$, $N_1$, and $N_2$, respectively. Then they are 
constrained by Euler's polyhedron formula,
\begin{eqnarray}
  \label{eq:epf}
  N_0-N_1+N_2=2.  
\end{eqnarray}
A regular polyhedron is specified by the Schl\"afli symbol $\{p,q\}$, 
where $p$ is the number of sides of each face and $q$ the number
of faces meeting at each vertex. This gives rise to the further 
constraints $N_0=2 N_1/q$ and $N_1=pN_2/2$. These, together 
with Eq. (\ref{eq:epf}), completely determine $N_{0,1,2}$.
In Table \ref{tab:rph} we summarize the properties of regular 
polyhedrons for the reader's reference. 

%%%%%%%%%%%%%%%%%%%%%%%%%%%%%%%%%%%%
%%%%%%%%%%%%%%  TABLE  %%%%%%%%%%%%%%%%
\begin{table}[t]
  \centering
  \begin{tabular}{cccccc}\hline 
    & Tetrahedron & ~~~~Cube~~~~ & ~Octahedron~ &  Dodecahedron 
    & Icosahedron \\ \hline
    $N_0$ & 4 & 8 & 6 & 20 & 12 \\
    $N_1$ & 6 & 12 & 12 & 30 & 30 \\
    $N_2$ & 4 & 6 & 8 & 12 & 20 \\ 
    $\{p,q\}$ & $\{3,3\}$ & $\{4,3\}$ & $\{3,4\}$ & $\{5,3\}$ 
    & $\{3,5\}$ \\ \hline
  \end{tabular}
  \caption{The five regular polyhedrons in three dimensions.}
  \label{tab:rph}
\end{table}
%%%%%%%%%%%%%%%%%%%%%%%%%%%%%%%%%%%
%%%%%%%%%%%%%%%%%%%%%%%%%%%%%%%%%%%%

The fundamental blocks of space-time in the Regge calculus are frustums 
with $p$-sided regular polygons as the upper and lower faces and 
$p$ isosceles trapezoids as the lateral faces, as depicted in 
Fig. \ref{fig:fbb}. We assume that the upper face of a frustum 
lies in a time-slice, so does the lower one. The whole space-time is 
then obtained by gluing such frustums face by face without a break. 
There are only two types of hinges in this piecewise linear manifold. 
One type of hinge is the edges of regular polyhedrons. We denote by 
$l_i$ the length of the edges on the $i$th Cauchy surface at time $t_i$. 
The other type is the struts between consecutive Cauchy surfaces. We 
denote by $m_i$ the length of the struts between the Cauchy surfaces at 
$t_i$ and $t_{i+1}$. Thus, the Regge action (\ref{eq:ract}) can be 
written as
\begin{eqnarray}
  \label{eq:regact}
  S_\mathrm{Regge}=\frac{1}{8\pi}\sum_i(N_0m_i\varepsilon_i^{\rm (s)}
  +N_1l_i\varepsilon_i^{\rm (e)}-N_2\Lambda V_i),
\end{eqnarray}
where $\varepsilon^{\rm (s)}_i$ and $\varepsilon_i^{\rm (e)}$ stand for 
the deficit angles around the strut and edge, respectively, and $V_i$ 
is the world-volume of the frustum. To avoid subtleties in defining 
lengths and angles, we assume for the time being the metric in 
each building block to be flat Euclidean, where geometric objects such 
as lengths and angles are obvious. The equations of motion in Lorentzian 
geometry can be achieved by the Wick rotation. 

%%%%%%%%%%%%%%%%%%%%%%%%%%%%%%%%%%%%
%%%%%%%%%%%%%%  FIGURE  %%%%%%%%%%%%%%%%
%%%%%%%%%%%%%%  FIGURE  %%%%%%%%%%%%%%%%
%%%%%%%%%%%%%%%%%%%%%%%%%%%%%%%%%%%%

%%%%%%%%%%%%%%%%%%%%%%%%%%%%%%%%%%%%%%%%%%%%
%%%%%%%%%%%%%%%%%%%%%%%%%%%%%%%%%%%%%%%%%%%%
%%%%%%%%%%%%%%%%%%%%%%%%%%%%%%%%%%%%%%%%%%%%

%                                << Sec.3 Regge equations >>

%%%%%%%%%%%%%%%%%%%%%%%%%%%%%%%%%%%%%%%%%%%%
%%%%%%%%%%%%%%%%%%%%%%%%%%%%%%%%%%%%%%%%%%%%
%%%%%%%%%%%%%%%%%%%%%%%%%%%%%%%%%%%%%%%%%%%%

\section{Regge equations}

\label{sec:req}
\setcounter{equation}{0}

The fundamental variables in Regge calculus are the lengths of the edges $l_i$  
and those of the struts $m_i$. The Regge equations can be obtained by applying 
the variational principle to the action (\ref{eq:regact}). Then, Eq. (\ref{eq:regeq}) 
can be written simply as 
\begin{align}
  \label{eq:regeqphuhc}
  \varepsilon_i^{(\mathrm{s})}&=\frac{q}{p}\Lambda\frac{\partial V_i}{\partial m_i}, \\
  \label{eq:regeqphumc}
  \varepsilon_i^{(\mathrm{e})}&=\frac{2}{p}\Lambda\left(\frac{\partial V_i}{\partial l_i}
    +\frac{\partial V_{i-1}}{\partial l_i}\right).
\end{align}
Note that the edge $l_i$ belongs to both $V_i$ and $V_{i-1}$. In the context 
of the ADM formalism, the first corresponds to the Hamiltonian constraint and the 
second to the evolution equation, respectively.

The volume and deficit angles can be expressed in terms 
of $l$'s and $m$'s. Since the inside of a frustum is flat Euclidean, 
we can compute the volumes and angles by standard geometric calculations.
The volume of the $i$th frustum is given by 
\begin{eqnarray}
  \label{eq:frvol}
  V_i=\frac{p}{12}(l^2_{i+1}+l_{i+1}l_i+l^2_i)
  \sqrt{m_i^2-\frac{1}{4}\delta l^2_i\csc^2\frac{\pi}{p}}\cot\frac{\pi}{p},
\end{eqnarray}
where we have introduced the variation of edge length $\delta l_i=l_{i+1}-l_i$. 

The deficit angle around a strut can be found by noting the fact that 
there are $q$ frustums having the strut in common. Then, 
$\varepsilon^{(\mathrm{s})}_i$ can be expressed as 
\begin{eqnarray}
  \label{eq:dasti}
  \varepsilon_i^{({\rm s})}=2\pi-q\theta_i, 
\end{eqnarray}
where $\theta_i$ is the dihedral angle between two adjacent lateral 
trapezoids. It is explicitly given by
\begin{eqnarray}
  \label{eq:dmi}
  \theta_i=\arccos\left(-\frac{4m_i^2\cos\frac{2\pi}{p}+\delta l_i^2}{%
      4m_i^2-\delta l_i^2}\right);
\end{eqnarray}
see Fig. \ref{fig:fbb}.  

To find the deficit angle around the edge $l_i$ we must take account of 
four frustums having the edge in common, two $V_i$ in the future side and 
two $V_{i-1}$ in the past side. Let $ \phi_i^{\uparrow}$ be the dihedral angle 
between the base regular polygon and a lateral trapezoid in the frustum 
$V_i$. Similarly, we denote by $\phi^\downarrow_i$ the dihedral angle 
between the top regular polygon and a lateral face in $V_{i-1}$ 
\cite{LW:2015aa,LW:2015ab}. 
As is easily seen in Fig. \ref{fig:fbb}, the dihedral angles are 
constrained by $\phi_i^{\uparrow}+\phi_{i+1}^{\downarrow}=\pi$. 
Then, the deficit angle $\varepsilon_i^{(\mathrm{e})}$ can be written as
\begin{eqnarray}
  \label{eq:daedg}
  \varepsilon_i^{({\rm e})}=2 \pi-2(\phi_i^{\uparrow}+\phi_i^{\downarrow})
  =2\delta\phi_i^{\downarrow},
\end{eqnarray}
where we have introduced $\delta\phi_i^{\downarrow}
=\phi_{i+1}^{\downarrow}-\phi_i^{\downarrow}$. In terms of the lengths of 
edges and struts, the dihedral angle $\phi^\downarrow_i$ can 
be expressed as
\begin{eqnarray}
  \label{eq:dhapda}
   \phi^\downarrow_i=\arccos\frac{\delta l_{i-1}\cot\frac{\pi}{p}}{%
      \sqrt{4m_{i-1}^2-\delta l_{i-1}^2}}.
\end{eqnarray}
Inserting these expressions for the deficit angles and volume element into 
the Regge equations (\ref{eq:regeqphuhc}) and (\ref{eq:regeqphumc}), 
we obtain a set of recurrence relations:
\begin{align}
  \label{eq:cfgvhc}
  2\pi-q\arccos\left(-\frac{4m_i^2\cos\frac{2\pi}{p}+\delta l_i^2}{%
      4m_i^2-\delta l_i^2}\right)=&\frac{q\Lambda}{12}
  \frac{(l^2_{i+1}+l_{i+1}l_i+l^2_i)m_i}{%
    \sqrt{m_i^2-\frac{1}{4}\delta l^2_i\csc^2\frac{\pi}{p}}}
  \cot\frac{\pi}{p}, \\
  \label{eq:cfgvmc}
  \arccos\frac{\delta l_i\cot\frac{\pi}{p}}{%
    \sqrt{4m_i^2-\delta l_i^2}}
  -\arccos\frac{\delta l_{i-1}\cot\frac{\pi}{p}}{%
    \sqrt{4m_{i-1}^2-\delta l_{i-1}^2}}
  =&\frac{\Lambda}{12}
  \Biggl[\frac{(l_{i+1}+2l_i)m_i^2
    +\frac{3}{4}l_i^2\delta l_i\csc^2\frac{\pi}{p}}{%
    \sqrt{m_i^2-\frac{1}{4}\delta l_i^2\csc^2\frac{\pi}{p}}} \nonumber\\
  &+\frac{(2l_i+l_{i-1})m_{i-1}^2
    -\frac{3}{4}l_i^2\delta l_{i-1}\csc^2\frac{\pi}{p}}{%
    \sqrt{m_{i-1}^2-\frac{1}{4}\delta l_{i-1}^2\csc^2\frac{\pi}{p}}}\Biggr]
  \cot\frac{\pi}{p}. \nonumber \\
\end{align}
As a consistency check, it is straightforward to see that 
these admit flat metric solutions for $\{p,q\}=\{3,6\}$, $\{4,4\}$, 
and $\{6,3\}$ in the absence of the cosmological term.

%%%%%%%%%%%%%%%%%%%%%%%%%%%%%%%%%%%%%%%%%%%%
%%%%%%%%%%%%%%%%%%%%%%%%%%%%%%%%%%%%%%%%%%%%
%%%%%%%%%%%%%%%%%%%%%%%%%%%%%%%%%%%%%%%%%%%%

%                                << Sec.4 Continuum time limit >>

%%%%%%%%%%%%%%%%%%%%%%%%%%%%%%%%%%%%%%%%%%%%
%%%%%%%%%%%%%%%%%%%%%%%%%%%%%%%%%%%%%%%%%%%%
%%%%%%%%%%%%%%%%%%%%%%%%%%%%%%%%%%%%%%%%%%%%

\section{Continuum time limit}

\label{sec:ctl}
\setcounter{equation}{0}

The nonlinear recurrence relations (\ref{eq:cfgvhc}) and 
(\ref{eq:cfgvmc}) are written only in terms of geometrical data, 
the edge and strut lengths $l_i$ and $m_i$. To obtain an insight into the 
evolution of the space-time we first note the relation between the 
strut length $m_i$ and the Euclidean time elapsed, 
$\delta t_i=t_{i+1}-t_i$. In Refs. \cite{LW:2015aa,LW:2015ab}, the 
time axis is chosen to be orthogonal to the Cauchy surface. For 
regular polyhedrons this works well. However, it does not work so well for 
general polyhedrons. To see this, let us consider two adjacent 
building blocks with equal height. If their base polygons are not 
congruent, two struts to be identified in gluing the building blocks 
would have different lengths. 
In the present polyhedral universe the spatial 
coordinates of each vertex are kept constant during the expansion. 
If we choose as $t_i$ the proper time of a clock expanding with a 
vertex of the polyhedral Cauchy surface, the strut length is 
given by 
\begin{eqnarray}
  \label{eq:misq}
  m_i=\delta t_i. 
\end{eqnarray}
This can be applied to any polyhedron. 

We assume, for simplicity, all the time intervals $\delta t_i$ to be equal 
and then take the continuum time limit $\delta t_i \to dt$. 
We can regard the edge length as a smooth function of time 
$l_i \to l(t)$, and 
\begin{eqnarray}
  \label{eq:ld}
  \delta l_i=\frac{\delta l_i}{\delta t_i}\delta t_i ~ \to ~
  \dot ldt, 
\end{eqnarray}
where $\dot l=dl/dt$. It is straightforward to compute the continuum 
time limit of Eqs. (\ref{eq:cfgvhc}) and (\ref{eq:cfgvmc}). We find:
\begin{align}
  \label{eq:chc}
  2\pi-q\arccos\frac{\dot l^2+4\cos\frac{2\pi}{p}}{%
      \dot l^2-4}
  &=\frac{q\Lambda}{2}\frac{l^2\cos\frac{\pi}{p}}{\sqrt{4\sin^2\frac{\pi}{p}-\dot l^2}}, \\
  \label{eq:cmc}
  \frac{\ddot l}{4-\dot l^2}&=-\frac{\Lambda}{4}l
  \left[1+\frac{l\ddot l}{2(4\sin^2\frac{\pi}{p}-\dot l^2)}\right].
\end{align}
Since we have fixed the strut lengths as mentioned above, they 
disappear from the Regge equations. One can easily 
verify that these are consistent each other. In other words, the 
Hamiltonian constraint can be obtained as the first integral of the 
evolution equation for the initial conditions \cite{LW:2015ab}:
\begin{eqnarray}
  \label{eq:24}
  l(0)=l_0= \sqrt{\frac{4\pi}{\Lambda}\left(\frac{2}{p}+\frac{2}{q}-1\right)
    \tan\frac{\pi}{p}}, \qquad \dot l(0)=0. 
\end{eqnarray}
The cosmological constant must be positive for regular 
polyhedrons. This implies that the space-time 
is de Sitter-like. The polyhedral universe cannot expand from or 
contract to a point but has minimum edge length $l_0$, as does the 
continuum solution, as we shall see below.

So far we have worked with Euclidean time. To discuss the evolution 
of space-time we move to the Minkowskian signature by Wick rotation. This 
can be done in Eqs. (\ref{eq:chc}) and (\ref{eq:cmc}) simply by letting
$\dot l^2, \ddot l \to -\dot l^2, -\ddot l$.
We thus obtain 
\begin{align}
  \label{eq:hcphu}
  2\pi-q\arccos\frac{\dot l^2-4\cos\frac{2\pi}{p}}{%
      4+\dot l^2}
  &=\frac{q\Lambda}{2}\frac{l^2\cos\frac{\pi}{p}}{\sqrt{4\sin^2\frac{\pi}{p}+\dot l^2}}, \\
  \label{eq:mcphu}
  \frac{\ddot l}{4+\dot l^2}&=\frac{\Lambda}{4}l
  \left[1-\frac{l\ddot l}{2(4\sin^2\frac{\pi}{p}+\dot l^2)}\right].
\end{align}
We can read off from 
the evolution equation that the acceleration $ \ddot{l} $ is always positive. 
Hence the universe expands as the continuum solution at the beginning for 
the initial conditions (\ref{eq:24}). The expansion, 
however, becomes much more rapid than the continuum solution as $t$ 
gets large, as we shall see in the next section.

%%%%%%%%%%%%%%%%%%%%%%%%%%%%%%%%%%%%%%%%%%%%
%%%%%%%%%%%%%%%%%%%%%%%%%%%%%%%%%%%%%%%%%%%%
%%%%%%%%%%%%%%%%%%%%%%%%%%%%%%%%%%%%%%%%%%%%

%                                << Sec.5 Numerical solution >>

%%%%%%%%%%%%%%%%%%%%%%%%%%%%%%%%%%%%%%%%%%%%
%%%%%%%%%%%%%%%%%%%%%%%%%%%%%%%%%%%%%%%%%%%%
%%%%%%%%%%%%%%%%%%%%%%%%%%%%%%%%%%%%%%%%%%%%

\section{Numerical solution}

\label{sec:nsol}
\setcounter{equation}{0}

The Hamiltonian constraint (\ref{eq:hcphu}) can be solved numerically. 
It is convenient to use the continuum limit of the dihedral angle $\theta_i$. 
Let us denote it by $\theta$; then $l$ and $\dot l$ can be expressed as 
\begin{align}
  \dot{l}^2&=4\sin^2\frac{\pi}{p}
  \left(\cot^2\frac{\pi}{p}\cot^2\frac{\theta}{2}-1\right), \label{eq:30} \\
  l^2&=\frac{4}{q\Lambda}\left(2\pi-q\theta\right)\cot\frac{\theta}{2}. \label{eq:31}
\end{align}
The first of these can be obtained directly from Eq. (\ref{eq:dmi}). The second 
can be derived from the Hamiltonian constraint (\ref{eq:chc}) by 
replacing $\dot l^2$ with Eq. (\ref{eq:30}). Since $\dot l^2\geq0$, the 
dihedral angle satisfies $0\leq\theta\leq\theta_p$, where $\theta_p=\frac{(p-2)\pi}{p}$ 
stands for the interior angle of a $p$-sided regular polygon. 
The edge length is a decreasing function of $\theta$ 
satisfying $l=l_0$ for $\theta=\theta_p$. As $\theta \to 0$, it 
approaches infinity. 

Eliminating the edge length from Eqs. (\ref{eq:30}) 
and (\ref{eq:31}), we obtain the differential equation describing the time development of 
$\theta$ as 
\begin{eqnarray}
  \dot{\theta}=\mp\frac{2\sqrt{2q\Lambda(2\pi-q\theta)\sin\theta
      \sin\displaystyle{\frac{\theta_p+\theta}{2}\sin\frac{\theta_p-\theta}{2}}}}{%
    2\pi-q(\theta-\sin\theta)}, \label{eq:33}
\end{eqnarray}
where the upper (lower) sign corresponds to an expanding (contracting) universe.  
As can be seen from Eqs. (\ref{eq:31}) and (\ref{eq:33}), the polyhedral universe expands to infinity 
at $t=t_{p,q}$ given by
\begin{eqnarray}
  \label{eq:tinfty}
  t_{p,q}=\int_0^{\theta_p}d\theta\frac{2\pi-q(\theta-\sin\theta)}{%
  2\sqrt{2q\Lambda(2\pi-q\theta)\sin\theta
      \sin\displaystyle{\frac{\theta_p+\theta}{2}\sin\frac{\theta_p-\theta}{2}}}}.
\end{eqnarray}

In what follows we focus our attention on the expanding universe.
Solving Eq. (\ref{eq:33}) numerically for the initial condition 
\begin{eqnarray}
\theta(0)=\theta_p-\epsilon, \label{eq:32}
\end{eqnarray}
we can find the evolution of the polyhedral universe, where we have 
introduced the positive infinitesimal $\epsilon$ to avoid the 
trivial solution $\theta(t)=\theta_p$. It is also possible to 
solve numerically the evolution equation (\ref{eq:mcphu}) 
with the initial condition (\ref{eq:24}). We shall use the latter 
approach in obtaining numerical solutions for the geodesic dome 
universe, where the edge lengths cannot be parametrized by a 
single dihedral angle. 

To compare with the continuum theory we must introduce an analog of 
the scale factor $a(t)$. In Ref. \cite{LW:2015ab} the authors 
discussed various definitions for the scale factor in a discretized 
FLRW universe in four dimensions. The behavior of universe, however, 
does not depend so much on the definition. Here, we simply define 
the scale factor of our polyhedral universe $a_\mathrm{R}$ as the 
radius of the circumsphere of the polyhedron. It is given by 
\begin{eqnarray}
  \label{eq:Rc}
  a_{\rm R}(t)=\frac{l(t)\sin\frac{\pi}{q}}{%
    2\sqrt{\sin^2\frac{\pi}{p}-\cos^2\frac{\pi}{q}}}.
\end{eqnarray}
The initial scale factor can be easily found as
\begin{eqnarray}
  a_{\rm R}(0)=\sqrt{\frac{\pi\left(\frac{2}{p}+\frac{2}{q}-1\right)
      \tan\frac{\pi}{p}}{\Lambda\left(\sin^2\frac{\pi}{p}-\cos^2\frac{\pi}{q}\right)}}
  \sin\frac{\pi}{q}. \label{eq:35} 
\end{eqnarray}

\begin{figure}[t]
  \centering
  \includegraphics[scale=0.8]{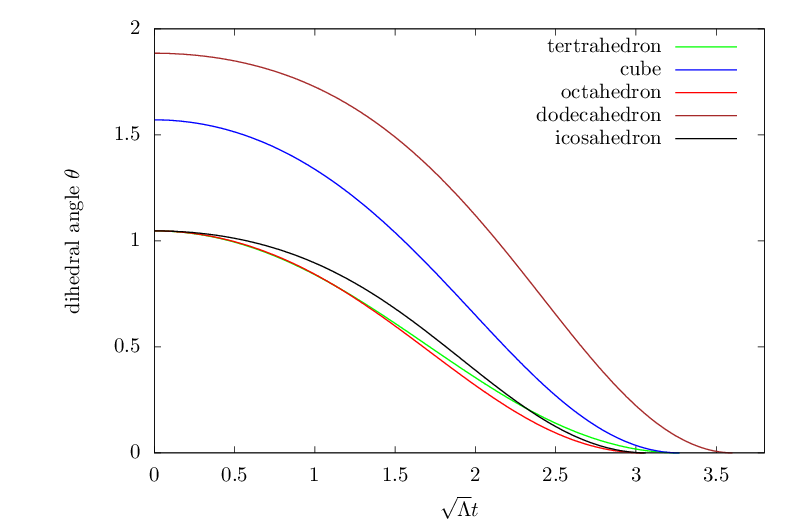}
  \caption{Plots of the dihedral angles of the regular polyhedral models: 
    each plot ends at $t=t_{p,q}$. }
  \label{fig:dhaphu}
\end{figure}

\begin{figure}[t]
  \centering
  \includegraphics[scale=0.8]{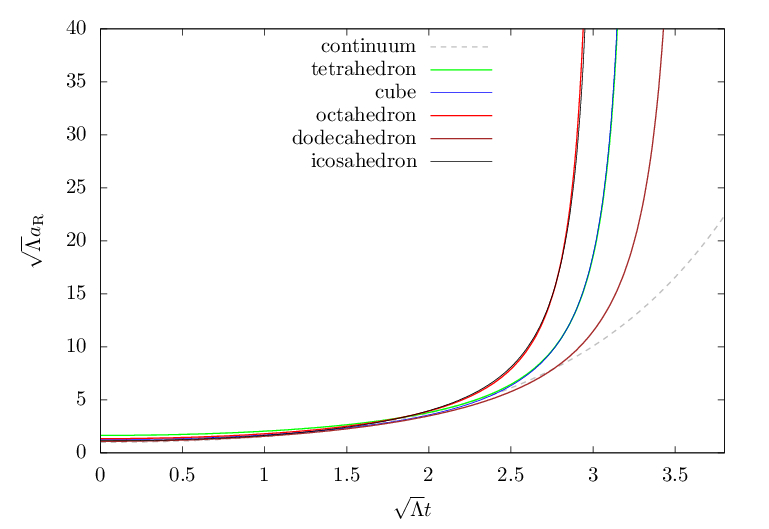}
  \caption{Plots of the scale factors of the regular polyhedral models.}
  \label{fig:sfrpu}
\end{figure}

In Fig. \ref{fig:dhaphu} the dihedral angles are plotted for the five 
types of regular polyhedrons. They are monotone decreasing functions 
of time and approach zero as $t \to t_{p,q}$. In Fig. \ref{fig:sfrpu}
we give the plots of the scale factors of the polyhedral universes as 
functions of time. The broken curve corresponds to the continuum solution. 
One can see that the polyhedral solutions approximate well the continuum at 
around $t=0$. This can be understood by noting the fact that the scale factor 
(\ref{eq:Rc}) approximately satisfies the Friedmann equation (\ref{eq:cc}) 
when both $\sqrt{\Lambda}l$ and $\dot l$ are small. 

The deviations from the continuum solution, however, get large for 
$t>2/\sqrt{\Lambda}$. The polyhedral universe expands much faster 
than the continuum FLRW universe and approaches an infinite 
size as $t \to t_{p,q}$. In fact, it is easy to see from Eqs. 
(\ref{eq:30}) and (\ref{eq:31}) that $a_\mathrm{R}(t)$ is approximately 
given by
\begin{eqnarray}
  \label{eq:asbaR}
  \sqrt{\Lambda}a_\mathrm{R}\approx
  \frac{c_{p,q}}{\sqrt{\Lambda}(t_{p,q}-t)},
\end{eqnarray}
where $c_{p,q}$ is defined by 
\begin{eqnarray}
  \label{eq:cpq}
  c_{p,q}=\frac{2\pi}{q}
  \frac{\sec\frac{\pi}{p}\sin\frac{\pi}{q}}{\sqrt{\sin^2\frac{\pi}{p}-\cos^2\frac{\pi}{q}}}.
\end{eqnarray} 
In Fig. \ref{fig:sfrpu} one can see the similarity between the octahedron 
and icosahedron. This can be understood from $c_{3,4}\approx c_{3,5}$ and $t_{3,4}
\approx t_{3,5}$. A similar thing also occurs for the tetrahedron and cube. 

%%%%%%%%%%%%%%%%%%%%%%%%%%%%%%%%%%%%
%%%%%%%%%%%%%%  FIGURE  %%%%%%%%%%%%%%%%
%%%%%%%%%%%%%%  FIGURE  %%%%%%%%%%%%%%%%
%%%%%%%%%%%%%%  FIGURE  %%%%%%%%%%%%%%%%
%%%%%%%%%%%%%%%%%%%%%%%%%%%%%%%%%%%%

%%%%%%%%%%%%%%%%%%%%%%%%%%%%%%%%%%%%%%%%%%%%
%%%%%%%%%%%%%%%%%%%%%%%%%%%%%%%%%%%%%%%%%%%%

%                                << Sec.6 Geodesic dome and pseudo-regular polyhedral universe >>

%%%%%%%%%%%%%%%%%%%%%%%%%%%%%%%%%%%%%%%%%%%%
%%%%%%%%%%%%%%%%%%%%%%%%%%%%%%%%%%%%%%%%%%%%
%%%%%%%%%%%%%%%%%%%%%%%%%%%%%%%%%%%%%%%%%%%%

\section{The geodesic dome and pseudo-regular polyhedral universes}

\label{sec:gdu}
\setcounter{equation}{0}

\begin{figure}[t]
  \centering
  \includegraphics[scale=0.6,angle=-90]{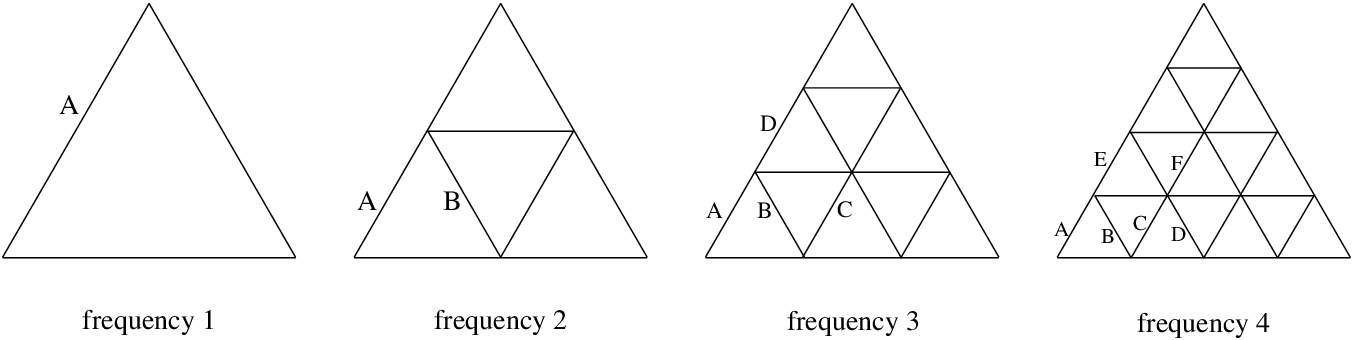}
  \caption{Subdivision of a regular triangle:  
    sides with distinct alphabetical labels are projected onto edges 
    of different lengths in the geodesic dome. }
  \label{fig:freq}
\end{figure}

\begin{figure}[t]
  \centering
  \begin{tabular}{cccc}
    \includegraphics[scale=.4]{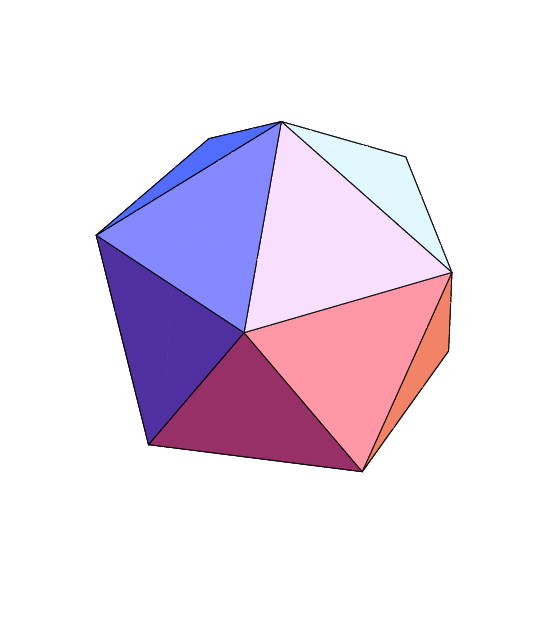} &
    \includegraphics[scale=.4]{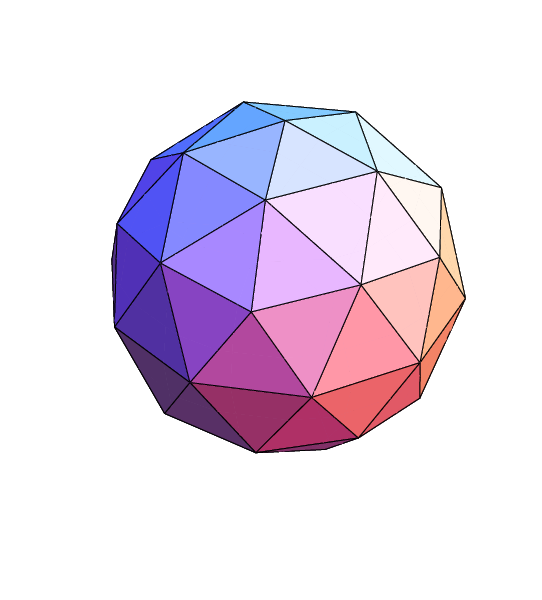} &
    \includegraphics[scale=.4]{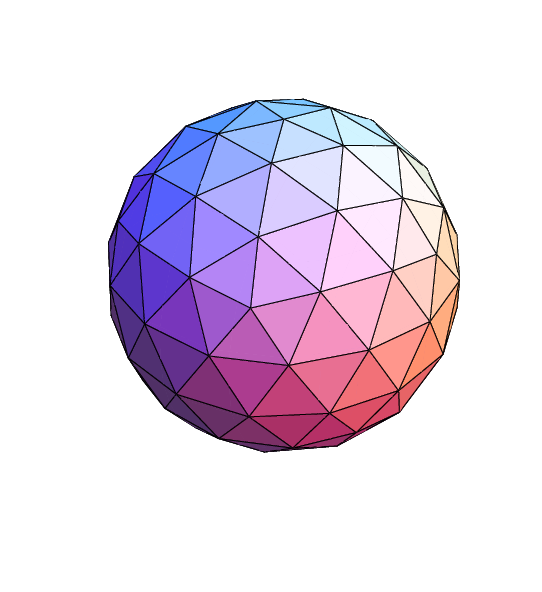} &
    \includegraphics[scale=.4]{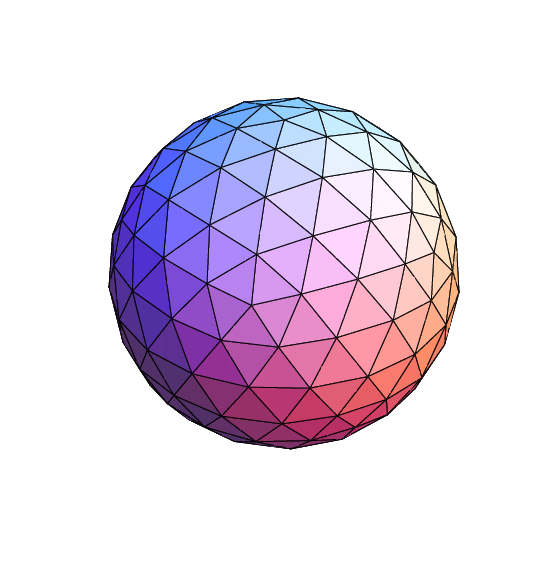}\\
    {\small frequency} 1 & {\small frequency 2}
    & {\small frequency 3} & {\small frequency 4}
  \end{tabular}
\caption{Projection of tessellated icosahedrons onto the circumsphere. 
}
\label{fig:gd}
\end{figure}

If we cease to stick to regular polyhedrons as the Cauchy surfaces, we can 
approximate a sphere more precisely. One way to put this into practice is to 
introduce geodesic domes. As depicted in Fig. \ref{fig:freq}, each face of 
a regular icosahedron can be subdivided into $\nu^2$ similar regular 
triangles, where $\nu=1,2,3, \cdots$ is the degree of subdivision, called 
the frequency. A geodesic dome is then obtained by projecting the icosahedron 
tessellated by the $20\nu^2$ triangular tiles onto the circumsphere. 
We show the first four geodesic domes in Fig. \ref{fig:gd}.   

%%%%%%%%%%%%%%%%%%%%%%%%%%%%%%%%%%%%
%%%%%%%%%%%%%%  FIGURE  %%%%%%%%%%%%%%%%
%%%%%%%%%%%%%%  FIGURE  %%%%%%%%%%%%%%%%
%%%%%%%%%%%%%%%  FIGURE  %%%%%%%%%%%%%%%%
%%%%%%%%%%%%%%%%%%%%%%%%%%%%%%%%%%%%%%%

The numbers of vertices, edges, and faces of the geodesic domes are summarized in 
Table \ref{tab:pgd}. Geodesic domes have only two types of connectors: 
the five-way connectors corresponding to the vertices of the original icosahedron, and 
the six-way connectors. Note that the larger the frequency, the more six-way connectors 
are used, while the number of five-way connectors is always 12. 
Furthermore, the faces of a geodesic dome are not regular triangles, except 
for the center triangles appearing for $\nu\equiv1,2\mod 3$. 
The number of types of edges with different lengths 
is given by 
\begin{eqnarray}
  \label{eq:ndte}
  M=
  \begin{cases}
    \displaystyle\frac{(\nu+1)^2}{4} & (\nu~\mathrm{odd}) , \\
    \displaystyle\frac{\nu^2+2\nu}{4} & (\nu~\mathrm{even}) .
  \end{cases}
\end{eqnarray}
Fortunately, we can take  as $l_i$ the length of an edge  between a 
five-way connector and a six-way one since any other edge lengths are 
proportional to $l_i$. Furthermore, all the struts connecting 
two adjacent Cauchy surfaces have the same length $m_i$ as in the 
case of the polyhedral universe. The geodesic dome model is then 
described by a single set of equations, the Hamiltonian constraint 
and the evolution equation. 
It is not difficult to carry out the Regge calculus for $\nu$ small,
as we show in the appendix. 
The number of different types of edges, however, grows roughly quadratically 
in $\nu/2$. This makes the Regge calculus rather cumbersome for large $\nu$.  

%%%%%%%%%%%%%%%%%%%%%%%%%%%%%%%%%%%%%%%
%%%%%%%%%%%%%%  TABLE  %%%%%%%%%%%%%%%%
\begin{table}[t]
  \centering
  \begin{tabular}{ccccccc}
    \hline & Frequency: & $1$ & $2$ & $3$ & $\ldots$ & $\nu$ \\
    \hline $N_2$ & & $20$ & $80$ & $180$ & $\ldots$ & $20 \nu^2$ \\
    \hline & Type A & $30$ & $60$ & $60$ & & \\
    & Type B & & $60$ & $90$ & & \\
    $N_1$ & Type C & & & $120$ & & \\
    & $\vdots$ &  &  & & & \\
    & Total & $30$ & $120$ & $270$ & $\ldots$ & $30 \nu^2$ \\
    \hline & Five-way connectors & $12$ & $12$ & $12$ & $\ldots$ & $12$ \\ 
    $N_0$ & Six-way connectors & $0$ & $30$ & $80$ & $\ldots$ & $10(\nu^2-1)$ \\
    & Total & $12$ & $42$ & $92$ & $\ldots$ & $10 \nu^2 + 2$ \\
    \hline
  \end{tabular}
  \caption{Numbers of faces, links, and vertices of geodesic domes.}
  \label{tab:pgd}
\end{table}
%%%%%%%%%%%%%%  TABLE  %%%%%%%%%%%%%%%%
%%%%%%%%%%%%%%%%%%%%%%%%%%%%%%%%%%%%%%%

To avoid the complexity of carrying out the Regge calculus for 
the geodesic dome, we regard it as a pseudo-regular polyhedron 
of edge length $l$ and assume that Eqs. (\ref{eq:hcphu}) and (\ref{eq:mcphu}) still hold 
true. How about the Schl\"afli symbol $\{p,q\}$? 
Since all the faces of the geodesic dome are triangles, we use $p=3$. As for $q$, 
we employ the average number of faces meeting at a vertex. We thus assign 
the fractional Schl\"afli symbol for the geodesic dome of frequency 
$\nu$ as
\begin{eqnarray}
  \label{eq:ssprp}
  \{p,q\}=\left\{3,\frac{60\nu^2}{10\nu^2+2}\right\}.
\end{eqnarray}
This recovers the icosahedron for $\nu=1$ and approaches $\{3,6\}$ 
as $\nu$ gets large, which corresponds to the tessellation of a flat 
plane with regular triangles. To see how the pseudo-regular polyhedron 
is close to the geodesic dome, we note that 
numerically the deviation of the averaged edge length $\bar l$ of the geodesic 
domes from $l$ satisfies, for $\nu\geq3$,
\begin{eqnarray}
  \label{eq:devlbl}
  \left|\frac{\bar l-l}{l}\right|<0.0013.
\end{eqnarray}
Hence it is legitimate to think of $l$ as the averaged edge length of the 
corresponding geodesic dome. 

If we introduce the dihedral angle for the pseudo-regular polyhedron 
universe by Eq. (\ref{eq:31}) as in the regular polyhedron cases 
and define the scale factor as Eq. (\ref{eq:Rc}), then 
the evolution of the universe can be described by Eq. (\ref{eq:33}) with 
the parameters (\ref{eq:ssprp}). The initial edge length is given by 
\begin{eqnarray}
  \label{eq:lzgdm}
  l(0)
  =\sqrt{\frac{4\sqrt{3}\pi}{\Lambda}\left(\frac{2}{q}-\frac{1}{3}\right)}
  =\frac{1}{\nu}\sqrt{\frac{4\sqrt{3}\pi}{15\Lambda}}. 
\end{eqnarray}
In contrast to the cases of regular polyhedrons,
this can be arbitrarily small as $\nu \to \infty$, or equivalently 
$q \to 6$. The initial scale factor (\ref{eq:35}), however, approaches 
the value of the continuum theory, as can easily be seen from 
\begin{eqnarray}
  \label{eq:isfgdm}
  a_\mathrm{R}(0)=2\sqrt{\frac{\sqrt{3}\pi(\frac{2}{q}-\frac{1}{3})}{%
      \Lambda(3-4\cos^2\frac{\pi}{q})}}
  \sin\frac{\pi}{q} ~\to~ \frac{1}{\sqrt{\Lambda}} \qquad
  (\nu ~ \to ~ \infty).
\end{eqnarray}

\begin{figure}[t]
  \centering
  \includegraphics[scale=0.8]{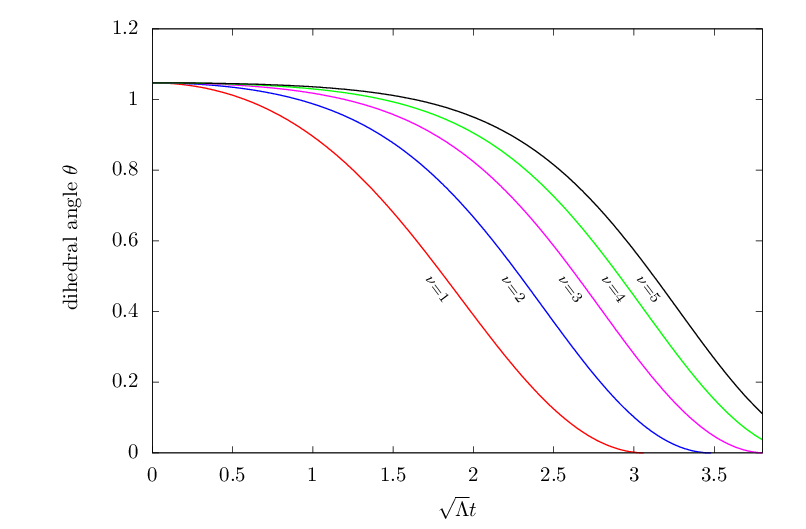}
  \caption{Plots of the dihedral angles of pseudo-regular polyhedral universes 
  for $\nu\leq5$.}
  \label{fig:dhagdm}
\end{figure}

\begin{figure}[t]
  \centering
  \includegraphics[scale=0.8]{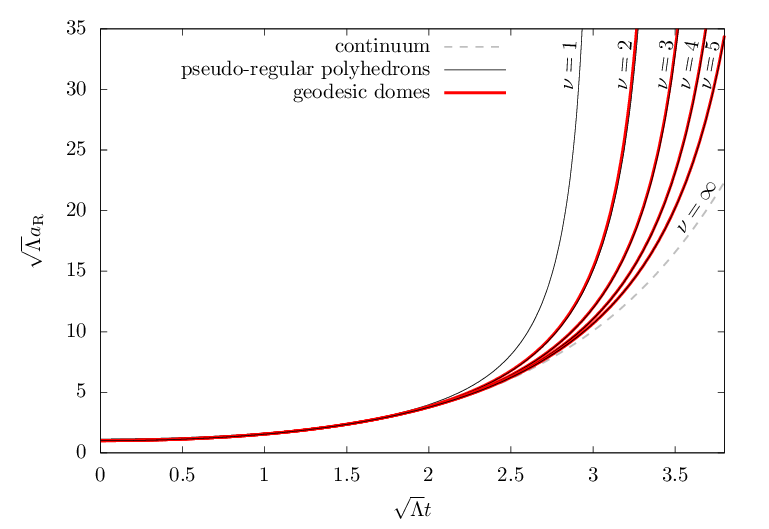}
  \caption{Plots of the scale factors of pseudo-regular polyhedrons 
    and geodesic domes for $\nu\leq5$.}
  \label{fig:sfgdm}
\end{figure}

There is no difficulty in numerically integrating (\ref{eq:33}) for fractional $q$. 
We give plots of the dihedral angle in Fig. \ref{fig:dhagdm} and the scale factor 
in Fig. \ref{fig:sfgdm} as functions of time for $\nu\leq5$. 
As mentioned before, the dihedral angle is a monotone decreasing 
function of time. The era of almost constant dihedral angle for small 
$\sqrt{\Lambda}t$ gets longer with the frequency. The time evolution of 
the dihedral angles becomes slower as we refine the triangulation of the Cauchy 
surface. The scale factor, however, develops as depicted in Fig. \ref{fig:sfgdm} 
since the ratio of the scale factor to the edge length becomes large
with the frequency. One easily sees that the results for the polyhedral universe 
given in Sect. \ref{sec:ctl} are improved more and more as $\nu$ increases. 

%%%%%%%%%%%%%%%%%%%%%%%%%%%%%%%%%%%%
%%%%%%%%%%%%%%  FIGURE  %%%%%%%%%%%%%%%%
%%%%%%%%%%%%%%  FIGURE  %%%%%%%%%%%%%%%%
%%%%%%%%%%%%%%%%%%%%%%%%%%%%%%%%%%%%

\begin{figure}[t]
  \centering
  \includegraphics[scale=0.8]{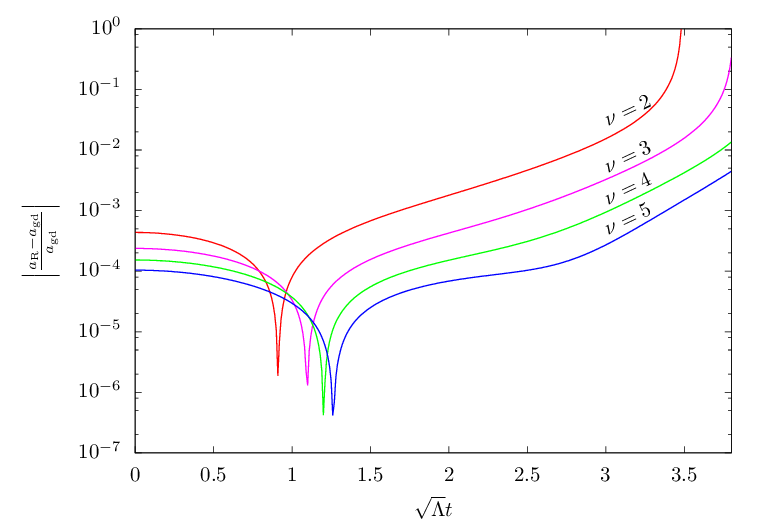}
  \caption{Plots of $|a_\mathrm{R}-a_\mathrm{gd}|/a_\mathrm{gd}$ for 
  $2\leq \nu\leq5$.}
  \label{fig:dev}
\end{figure}

As mentioned above, the pseudo-regular polyhedron is not the geodesic dome. 
Readers may wonder to what extent our model reproduces the geodesic 
dome universe. It is possible to justify our approach by comparing 
with the results of the Regge calculus for the geodesic dome universe. We 
have carried this out for $\nu\leq5$. To give readers 
some feeling for the Regge calculus, the Hamiltonian 
constraints for the geodesic dome universes with $\nu=2$ and $\nu=3$ are 
shown in the appendix. The evolution equations can be obtained by differentiating 
them with respect to time. We can also read off the initial conditions from 
the Hamiltonian constraints. It is straightforward to obtain numerical 
solutions to the evolution equations. The results 
are plotted in Fig. \ref{fig:sfgdm}. The plots for the pseudo-regular 
polyhedrons almost overlap with those for the corresponding geodesic domes. 
The rates of deviation of the scale factor $a_\mathrm{R}$ 
for the pseudo-regular polyhedron universe from the scale factor 
$a_\mathrm{gd}$ of the corresponding geodesic dome universe 
can be found in Fig. \ref{fig:dev}.   
We see that the pseudo-regular polyhedron model better approximates the 
geodesic dome universe as the frequency increases. 

\begin{figure}[t]
  \centering
  \includegraphics[scale=0.8]{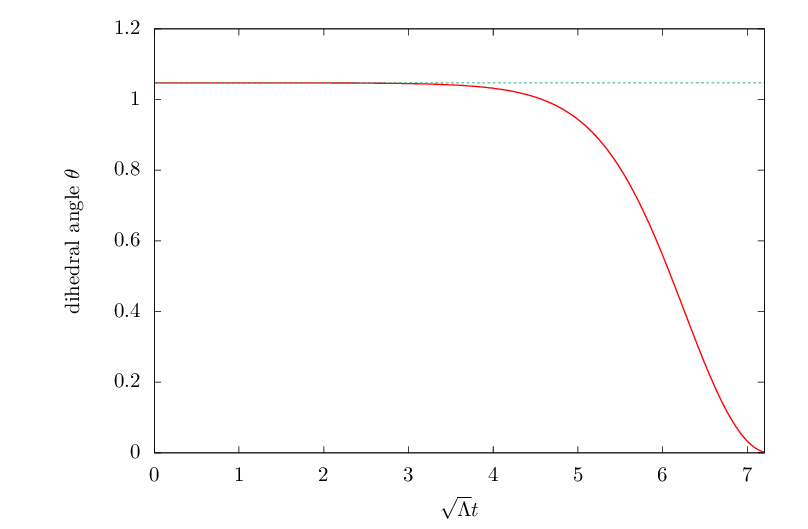}
  \caption{Dihedral angle of the pseudo-regular polyhedron for $\nu=100$. The broken 
  line corresponds to the solution in the infinite frequency limit $\theta_\infty(t)=\pi/3$.}
  \label{fig:lflthgdm}
\end{figure}

\begin{figure}[t]
  \centering
  \includegraphics[scale=0.8]{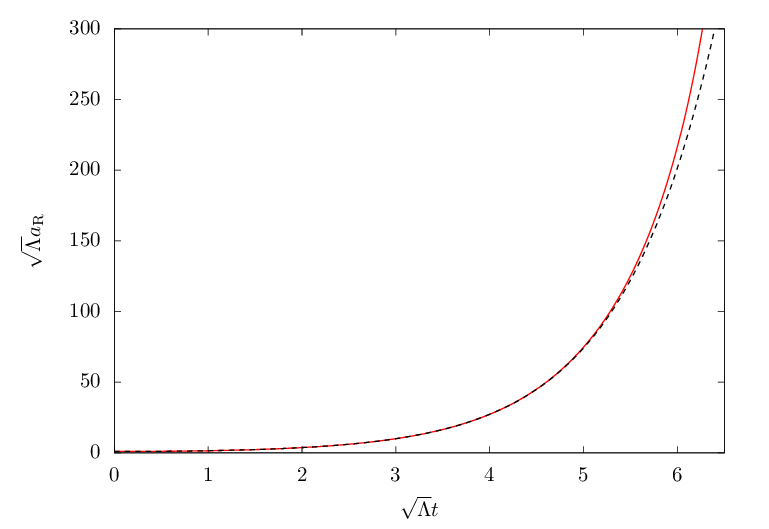}
  \caption{Scale factor of the pseudo-regular polyhedron for $\nu=100$. The broken curve 
  is the scale factor of the continuum theory.}
  \label{fig:lflargdm}
\end{figure}

Having established the relation with the geodesic dome universe, we 
can apply the pseudo-regular polyhedron model to the cases where 
the direct Regge calculus can hardly be practical. 
In Fig. \ref{fig:lflthgdm} we give the dihedral angle for $\nu=100$. 
It approaches a constant solution,
\begin{eqnarray}
  \label{eq:dhaiifl}
  \theta(t) ~ \to ~ \theta_\infty(t)=\frac{\pi}{3},
\end{eqnarray}
in the infinite frequency limit. Indeed, $\theta_\infty(t)$ satisfies 
Eqs. (\ref{eq:32}) and (\ref{eq:33}), as one can verify directly. The scale 
factor for the pseudo-regular polyhedral universe with $\nu=100$ is plotted in 
Fig. \ref{fig:lflargdm}. During the era of almost constant dihedral 
angle the agreement with the continuum scale factor can be seen 
immediately. 

%%%%%%%%%%%%%%%%%%%%%%%%%%%%%%%%%%%%
%%%%%%%%%%%%%%  FIGURE  %%%%%%%%%%%%%%%%
%%%%%%%%%%%%%%  FIGURE  %%%%%%%%%%%%%%%%
%%%%%%%%%%%%%%  FIGURE  %%%%%%%%%%%%%%%%
%%%%%%%%%%%%%%  FIGURE  %%%%%%%%%%%%%%%%
%%%%%%%%%%%%%%  FIGURE  %%%%%%%%%%%%%%%%
%%%%%%%%%%%%%%%%%%%%%%%%%%%%%%%%%%%%

Finally, it can be shown that the scale factor of the pseudo-regular polyhedral universe,
\begin{eqnarray}
  \label{eq:sfgdu}
  a_\mathrm{R}(t)=\frac{l(t)\sin\frac{\pi}{q}}{\sqrt{3-4\cos^2\frac{\pi}{q}}},
\end{eqnarray}
approaches that of the continuum theory in the infinite frequency 
limit. The easiest way to see this is to come back to Eqs. (\ref{eq:hcphu}) 
and (\ref{eq:mcphu}). Rewriting them in terms of $a_\mathrm{R}$ and then 
taking the limit $q \to 6$, we obtain the continuum equations 
of the scale factor (\ref{eq:cc}). This also justifies the approach 
regarding Eqs. (\ref{eq:hcphu}) and  (\ref{eq:mcphu}) as the equations 
of motion of the discretized FLRW universe.

%%%%%%%%%%%%%%%%%%%%%%%%%%%%%%%%%%%%%%%%%%%%
%%%%%%%%%%%%%%%%%%%%%%%%%%%%%%%%%%%%%%%%%%%%
%%%%%%%%%%%%%%%%%%%%%%%%%%%%%%%%%%%%%%%%%%%%

%   << Sec.7 Summary and discussions >>

%%%%%%%%%%%%%%%%%%%%%%%%%%%%%%%%%%%%%%%%%%%%
%%%%%%%%%%%%%%%%%%%%%%%%%%%%%%%%%%%%%%%%%%%%
%%%%%%%%%%%%%%%%%%%%%%%%%%%%%%%%%%%%%%%%%%%%

\section{Summary and discussions}

\label{sec:sum}
\setcounter{equation}{0}

We have investigated the closed FLRW universe in three dimensions 
using the CW formalism in Regge calculus. We have given 
unified expressions applicable to any regular polyhedron as 
the Cauchy surfaces. We have shown that the Regge equations in 
the continuum Lorentzian time limit, one corresponding to 
the Hamiltonian constraint and the other to the evolution equation, 
describe the evolution of the universe. In spite of the simplest 
approximations of spheres by regular polyhedrons, the 
coincidence with the continuum solution is 
appreciable when the size of the universe is around 
the minimum, where the nonlinearity of the evolution 
equation (\ref{eq:mcphu}) can be neglected. The discrepancies, 
however, become larger and larger with the size of the universe. 
The polyhedral universe expands much faster than the continuum 
FLRW universe. This is because the nonlinearity always 
enhances the acceleration of the universe. 

The expansion of the regular polyhedral universe can be slowed 
down by refining the triangulation of the Cauchy surface. 
To carry this out systematically we employed geodesic dome 
models and introduced pseudo-regular polyhedrons. 
We proposed the pseudo-regular polyhedral universe 
described by Eqs. (\ref{eq:hcphu}) and (\ref{eq:mcphu}) 
with the fractional Schl\"afli symbol (\ref{eq:ssprp}) as a 
substitute for geodesic domes. We have shown that our 
model considerably approximates the geodesic dome universes. 
As it should be, the continuum solution can be obtained in 
the infinite frequency limit. 

We have considered a closed compact universe. In the continuum 
theory there is an oscillating solution for a negative cosmological 
constant, where the Cauchy surface is not compact. Application 
of our approach to a hyperspherical universe might be interesting. 
Our concern in this work was the vacuum solution. Hence, the inclusion 
of matter would be worth investigation. We have worked with three dimensions. 
It would be of great interest to extend our approach to the four-dimensional 
FLRW universe. 

\vskip .5cm

%%%%%%%%%%%%%%%%%%%%%%%%%%%%%%%%%%%%%%%%%%%%
%%%%%%%%%%%%%%%%%%%%%%%%%%%%%%%%%%%%%%%%%%%%
%%%%%%%%%%%%%%%%%%%%%%%%%%%%%%%%%%%%%%%%%%%%

%                                << acknowledgments>>

%%%%%%%%%%%%%%%%%%%%%%%%%%%%%%%%%%%%%%%%%%%%
%%%%%%%%%%%%%%%%%%%%%%%%%%%%%%%%%%%%%%%%%%%%
%%%%%%%%%%%%%%%%%%%%%%%%%%%%%%%%%%%%%%%%%%%%

\begin{center}
  {\bf Acknowledgments}
\end{center}
The authors would like to thank Y. Hyakutake, N. Motoyui, 
M. Sakaguchi, and S. Tomizawa for useful discussions. This
work was supported in part by Grant-in-Aid for
Scientific Research Number 24540247 from the Japan Society 
for the Promotion of Science.

%%%%%%%%%%%%%%%%%%%%%%%%%%%%%%%%%%%%%%%%%%%%
%%%%%%%%%%%%%%%%%%%%%%%%%%%%%%%%%%%%%%%%%%%%
%%%%%%%%%%%%%%%%%%%%%%%%%%%%%%%%%%%%%%%%%%%%

%                                << Appendix Regge calculus for the geodesic domes >>

%%%%%%%%%%%%%%%%%%%%%%%%%%%%%%%%%%%%%%%%%%%%
%%%%%%%%%%%%%%%%%%%%%%%%%%%%%%%%%%%%%%%%%%%%
%%%%%%%%%%%%%%%%%%%%%%%%%%%%%%%%%%%%%%%%%%%%

\appendix

\section{Regge calculus for the geodesic domes}

\label{sec:hcgd}

%%%%%%%%%%%%%%%%%%%%%%%%%%%%%%%%%%%%
%%%%%%%%%%%%%%  FIGURE  %%%%%%%%%%%%%%%%
\begin{figure}[t]
  \centering
  \includegraphics[scale=1.0]{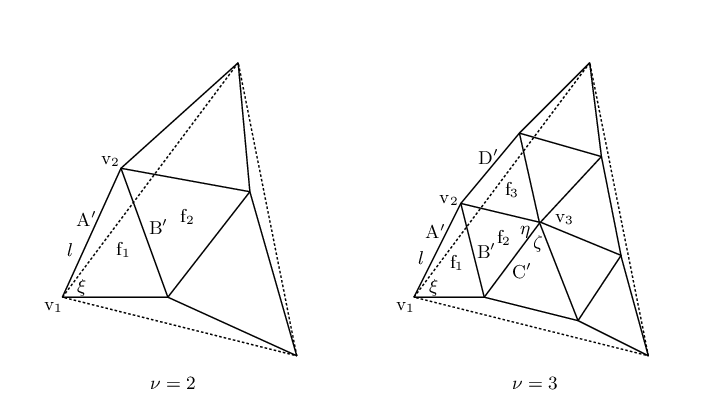}
  \caption{Faces projected onto the sphere.}
  \label{fig:gdnu23}
\end{figure}
%%%%%%%%%%%%%%  FIGURE  %%%%%%%%%%%%%%%%
%%%%%%%%%%%%%%%%%%%%%%%%%%%%%%%%%%%%

In this appendix we give the Hamiltonian constraints for the geodesic dome 
models of frequencies $\nu=2$ and $3$. The subdivided faces in Fig. 
\ref{fig:freq} are projected onto the sphere as depicted in Fig.
\ref{fig:gdnu23}. Edges of different lengths are labeled A$^\prime$, 
B$^\prime$, $\ldots$ 
Faces and vertices of different types are denoted f$_n$ and v$_m$, 
respectively. We choose the length of type A$^\prime$ edge as $l_i$. 
All other edge lengths can be expressed by $l_i$ and the angles 
$\xi$, $\eta$, and $\zeta$. 

To write down the Hamiltonian constraint we need the deficit angle 
$\varepsilon^{(\mathrm{s})}_{m,i}$ about the strut $m_i$ at the vertex v$_m$ 
and the volume element $V_{n,i}$ with f$_n$ as the bottom face. 
The deficit angles can be expressed by dihedral angles as Eq.
(\ref{eq:dasti}). Since a dihedral angle appearing in $V_{n,i}$ 
is uniquely determined by specifying a strut in the volume element, we denote 
the dihedral angle about the strut $m_i$ at the vertex v$_m$ 
by $\theta_{m,n,i}$. Then the deficit angle 
$\varepsilon^{(\mathrm{s})}_{m,i}$ is given by
\begin{eqnarray}
  \label{eq:dami}
  \varepsilon^{(\mathrm{s})}_{m,i}=2\pi -\sum_n\theta_{m,n,i},
\end{eqnarray}
where the summation must be taken over the indices $n$ of the volume 
elements with strut $m_i$ in common. There are three different types of dihedral 
angles for $\nu=2$ case: $\theta_{1,1,i}$, $\theta_{2,1,i}$, 
and $\theta_{2,2,i}$. The number of independent dihedral angles 
is six for $\nu=3$. Concrete expressions for $V_{n,i}$ and
$\theta_{m,n,i}$ in terms of $l_i$ and $m_i$ can be found by
standard geometry. 

We are now able to write the Hamiltonian constraints. They are:
\begin{align}
  \label{eq:nu2hc}
  \nu=2&:\frac{1}{5}\varepsilon^{(\mathrm{s})}_{1,i}
  +\frac{1}{2}\varepsilon^{(\mathrm{s})}_{2,i}
  =\Lambda\left(\frac{\partial V_{1,i}}{\partial m_i}
    +\frac{1}{3}\frac{\partial V_{2,i}}{\partial m_i}\right), \\
  \label{eq:nu3hc}
  \nu=3&:\frac{1}{5}\varepsilon^{(\mathrm{s})}_{1,i}
  +\varepsilon^{(\mathrm{s})}_{2,i}
  +\frac{1}{3}\varepsilon^{(\mathrm{s})}_{3,i}
  =\Lambda\left(\frac{\partial V_{1,i}}{\partial m_i}
    +\frac{\partial V_{2,i}}{\partial m_i}
    +\frac{\partial V_{3,i}}{\partial m_i}\right).
\end{align}
The coefficients in front of the deficit angles and volume elements
can be found by counting the number of vertices of type v$_m$ 
and the number of volume elements with f$_n$ as the bottom face. 

Taking the limit of continuum time followed by the Wick rotation, we 
finally obtain, for $\nu=2$,
\begin{eqnarray}
  \label{eq:hcnu2}
  \frac{1}{5}\varepsilon^{(\mathrm{s})}_{1}
  +\frac{1}{2}\varepsilon^{(\mathrm{s})}_{2}
  =\Lambda l^2\left[\frac{\sin\xi\cos\frac{\xi}{2}}{%
      \sqrt{4\cos^2\frac{\xi}{2}+\dot l^2}}
    +\frac{\sin^2\frac{\xi}{2}}{%
      \sqrt{3+4\dot l^2\sin^2\frac{\xi}{2}}}\right] .
\end{eqnarray}
The deficit angles are given by
\begin{align}
  \label{eq:danu2}
  \varepsilon^{(\mathrm{s})}_{1}&=2\pi-5\theta_{1,1},  \\
  \varepsilon^{(\mathrm{s})}_{2}&=2\pi-4\theta_{2,1}-2\theta_{2,2},
\end{align}
where the dihedral angles are 
\begin{align}
  \label{eq:dhanu2}
  \theta_{1,1}&=\arccos\frac{4\cos\xi+\dot l^2}{4+\dot l^2}, \\
  \theta_{2,1}&=\arccos\frac{(2+\dot l^2)\sin\frac{\xi}{2}}{%
    \sqrt{(4+\dot l^2)(1+\dot l^2\sin^2\frac{\xi}{2})}}, \\
  \theta_{2,2}&=\arccos\frac{1+2\dot l^2\sin^2\frac{\xi}{2}}{%
    2(1+\dot l^2\sin^2\frac{\xi}{2})}.
\end{align}
For $\nu=3$ the deficit angles can be found as
\begin{align}
  \label{eq:danu3}
  \varepsilon^{(\mathrm{s})}_{1}&=2\pi-5\theta_{1,1},  \\
  \varepsilon^{(\mathrm{s})}_{2}&=2\pi-2(\theta_{2,1}+\theta_{2,2}+\theta_{2,3}), \\
  \varepsilon^{(\mathrm{s})}_{3}&=2\pi-3(\theta_{3,2}+\theta_{3,3}),
\end{align}
with 
\begin{align}
  \label{eq:dhanu3}
  \theta_{1,1}&=\arccos\frac{4\cos\xi+\dot l^2}{4+\dot l^2}, \\
  \theta_{2,1}&=\arccos\frac{(2+\dot l^2)\sin\frac{\xi}{2}}{%
    \sqrt{(4+\dot l^2)(1+\dot l^2\sin^2\frac{\xi}{2})}}, \\
  \theta_{2,2}&=\arccos\frac{2\sin^2\frac{\eta}{2}+\dot l^2\sin^2\frac{\xi}{2}}{%
    \sqrt{(1+\dot l^2\sin^2\frac{\xi}{2})
      (4\sin^2\frac{\eta}{2}+\dot l^2\sin^2\frac{\xi}{2})}}, \\
  \theta_{2,3}&=\arccos\frac{(2\sin^2\frac{\eta}{2}
    +\dot l^2\sin^2\frac{\xi}{2})\sin\frac{\zeta}{2}}{%
    \sqrt{(4\sin^2\frac{\eta}{2}+\dot l^2\sin^2\frac{\xi}{2})
      (\sin^2\frac{\eta}{2}+\dot l^2\sin^2\frac{\xi}{2}\sin^2\frac{\zeta}{2})}}, \\
  \theta_{3,2}&=\arccos\frac{4\cos\eta\sin^2\frac{\eta}{2}+\dot l^2\sin^2\frac{\xi}{2}}{%
    4\sin^2\frac{\eta}{2}+\dot l^2\sin^2\frac{\xi}{2}}, \\
  \theta_{3,3}&=\arccos\frac{4\cos\zeta\sin^2\frac{\eta}{2}+\dot l^2\sin^2\frac{\xi}{2}}{%
    4\sin^2\frac{\eta}{2}+\dot l^2\sin^2\frac{\xi}{2}}.
\end{align}
The Hamiltonian constraint is then given by 
\begin{align}
  \label{eq:hcnu3}
  & \frac{1}{5}\varepsilon^{(\mathrm{s})}_{1}
  +\varepsilon^{(\mathrm{s})}_{2}
  +\frac{1}{3}\varepsilon^{(\mathrm{s})}_{3} \nonumber \\
  &=\Lambda l^2\left[\frac{\sin\xi\cos\frac{\xi}{2}}{\sqrt{4\cos^2\frac{\xi}{2}+\dot l^2}}
    +\frac{2\sin^2\frac{\xi}{2}\cos^2\frac{\eta}{2}}{%
      \sqrt{\sin^2\eta+\dot l^2\sin^2\frac{\xi}{2}}}
    +\frac{\sin^2\frac{\xi}{2}\csc\frac{\eta}{2}\sin\zeta\cos\frac{\zeta}{2}}{%
      \sqrt{4\sin^2\frac{\eta}{2}\cos^2\frac{\zeta}{2}+\dot l^2\sin^2\frac{\xi}{2}}}
    \right]. 
\end{align}
The evolution equations can be obtained similarly. They can also be
obtained by taking the time derivative of the Hamiltonian constraints. 
Since they are linear with respect to the second derivative $\ddot l$, 
numerical integration is straightforward. In doing this we need 
initial conditions. We can assume $\dot l(0)=0$. The initial edge 
length $l(0)$ can be found from the Hamiltonian constraints. Finally, 
we define the scale factor $a_\mathrm{gd}$ for the geodesic dome by the 
radius of the circumsphere. It is nothing but the radius of the circumsphere
of the parent icosahedron of the geodesic dome.

%%%%%%%%%%%%%%%%%%%%%%%%%%%%%%%%%%%%%%%%%%%%
%%%%%%%%%%%%%%%%%%%%%%%%%%%%%%%%%%%%%%%%%%%%
%%%%%%%%%%%%%%%%%%%%%%%%%%%%%%%%%%%%%%%%%%%%

%                                << Reference>>

%%%%%%%%%%%%%%%%%%%%%%%%%%%%%%%%%%%%%%%%%%%%
%%%%%%%%%%%%%%%%%%%%%%%%%%%%%%%%%%%%%%%%%%%%
%%%%%%%%%%%%%%%%%%%%%%%%%%%%%%%%%%%%%%%%%%%%

\end{document}